\documentstyle[aas2pp4]{article}

\begin{document}

\title{Viscosity in Selfgravitating Accretion Disks}
\author{Wolfgang J.\ Duschl \altaffilmark{1}}
\affil{Institut f\"ur Theoretische Astrophysik, Tiergartenstr.\ 15,
D-69121 Heidelberg, Germany}
\authoremail{wjd@ita.uni-heidelberg.de}
\author{Peter A.\ Strittmatter \altaffilmark{1}}
\affil{Steward Observatory, The University of Arizona, Tucson, AZ 85721, USA}
\authoremail{pstrittmatter@as.arizona.edu}
\and
\author{Peter L.\ Biermann}
\affil{Max-Planck-Institut f\"ur Radioastronomie, Auf dem H\"ugel 69,
D-53121 Bonn, Germany}
\authoremail{p165bie@mpifr-bonn.mpg.de}
\altaffiltext{1}{also at: Max-Planck-Institut f\"ur Radioastronomie, Bonn,
Germany}

\begin{abstract}
We show that the standard model for geometrically thin accretion disks
($\alpha$-disks) leads to inconsistencies if selfgravity plays a role.
This problem arises from the parametrization of viscosity in terms of
local sound velocity and vertical disk scale height. A viscosity prescription
based on turbulent flows at the critical effective Reynolds number allows 
for consistent models of thin selfgravitating disks, and recovers the
$\alpha$-disk solution as the limiting case of negligible selfgravity. We 
suggest that such selfgravitating disks may explain the observed spectra of 
protoplanetary disks and yield a natural explanation for the radial motions 
inferred from the observed metallicity gradients in disk galaxies.
\end{abstract}


\section{Introduction\label{introduction}}

One of the major shortcomings of the current theoretical descriptions of
accretion disks is lack of detailed knowledge about the underlying physics
of viscosity in the disk. This problem is significant because almost all
detailed modelling of the structure and evolution of accretion disks depends
on the value of the viscosity and its dependence on the physical parameters.
While there is general agreement that molecular viscosity is totally
inadequate and that some kind of turbulent viscosity is required, there is
far less certainty about its prescription. Most investigators adopt the
so-called $\alpha$-{\it ansatz\/} introduced by Shakura (1972) and 
Shakura and Sunyaev (1973)
that gives the viscosity ($\nu$) as the product of the pressure scale height
in the disk ($h$), the velocity of sound ($c_{{\rm s}}$), and a parameter 
$\alpha$ that contains all the unknown physics. One interprets this as some
kind of isotropic turbulent viscosity $\nu = \nu_{{\rm t}} = l_{{\rm t}} 
v_{{\rm t}}$ where $l_{{\rm t}}$ is an (unknown) length scale and $v_{{\rm t}}$
an (unknown) characteristic velocity of the turbulence. One may then write 
$\alpha = (v_{{\rm t}}/c_{{\rm s}})\cdot(l_{{\rm t}}/h)$. On general physical
grounds neither term in parentheses can exceed unity so that $\alpha \le 1$.
If initially $v_{{\rm t}} > c_{{\rm s}}$, shock waves would result in strong
damping and hence a return to a subsonic turbulent velocity. The condition 
$l_{{\rm t}} > h$ would require anisotropic turbulence since the vertical
length scales are limited by the disk's thickness, which is comparable to $h$.

A parameterization of this sort for $\nu$ is of course only of interest if
the proportionality parameter, $\alpha$, is (approximately) a constant. One
can expect this to happen only if the scaling quantities are chosen in a
physically appropriate manner. Models for the structure and evolution of
accretion disks in close binary systems (e.g., dwarf novae and symbiotic
stars) show that Shakura and Sunyaev's parameterization with a constant 
$\alpha$ leads to results that reproduce the overall observed behaviour of
the disks quite well. As a result of this success, the $\alpha$-ansatz is
now used in practice in all sorts of accretion disks. It is noteworthy,
however, that the $\alpha$-ansatz retains no information about the
mechanism generating the turbulence but only about limits to its efficiency.

In this contribution, we wish to

\begin{itemize}
\item  show that simple application of the above viscosity prescription to
the case of selfgravitating accretion disks leads to an apparent internal
inconsistency (Sec.\ \ref{inconsistency}).

\item  propose a possible resolution of this problem in terms of a viscosity
prescription (Sec.\ \ref{solutioneins}).

\item  discuss the implications of the new formulation of viscosity for the
structure and evolution of accretion disks (Sec.\ \ref{structure}).

\item  discuss protostellar and galactic disks as two examples where a
non-standard  viscosity description in selfgravitating disks is important
(Sec.\ \ref{examples}).
\end{itemize}

\section{The Structure of Selfgravitating Accretion Disks}

\subsection{Conditions for Selfgravity in Accretion Disks}

In the following we assume that the accretion disks are geometrically thin
in the vertical direction and symmetric in the 
azimuthal direction. We approximate
the vertical structure by a one zone model. Then a disk model is specified
by the central mass $M_*$, the radial distributions of surface density 
$\Sigma(s)$, central plane temperature $T_{{\rm c}}(s)$ and effective
temperature $T_{{\rm eff}}(s)$. The relevant material functions are the
equation of state, the opacity and the viscosity prescription.

One can estimate the importance of selfgravity by comparing the respective
contributions to the local gravitational accelerations in the vertical and
radial directions.

The vertical gravitational acceleration at the disk surface is $2 \pi G \Sigma$
and $G M_* h / s^3$, respectively, for the selfgravitating and the purely
Keplerian case, respectively. Selfgravitation is thus dominant in the
vertical direction when

\begin{equation}
\label{eqcondvertsg}
\frac{M_{\rm d}}{M_*} \sim \frac{\pi s^2 \Sigma}{M_*} > \frac{1}{2} 
\frac{h}{s},
\end{equation}
where $M_{\rm d}(s)$ is the enclosed mass in the disk and is given 
approximately by $M_{\rm d} \sim \pi s^2 \Sigma$.
With characteristic numbers for the relative geometrical thickness of
non-selfgravitating disks of $h/s \sim 0.1\ \dots\ 0.02$, condition
(\ref{eqcondvertsg}) translates into a condition for vertical selfgravity
of $M_{\rm d}/M_* > 0.01\ \dots\ 0.05$.

Similar considerations lead to the condition

\begin{equation}
\label{eqcondradsg}
M_{\rm d} > M_*
\end{equation}
for selfgravitation to dominate in the radial direction.
Thus, for increasing disk masses selfgravity first becomes important in 
the vertical direction.

\subsection{Keplerian Selfgravitating Disks (KSG)\label{inconsistency}}

In this Section, we describe a selfgravitating (SG) accretion disk in the
framework of the model by Shakura and Sunyaev (1973; hereafter referred to as 
the {\it standard model\/}), changing only the equation of hydrostatic
equilibrium in the direction perpendicular to the disk's plane ({\it 
vertical\/} or $z$ direction) to account for selfgravity in the direction 
perpendicular to the disk. Thus, while in the standard model the local vertical
pressure gradient is balanced by the $z$ component of the gravitational
force due to the central object, in the SG case we have balance between two
local forces, namely the pressure force and the gravitational force due to
the disk's local mass. For the present purpose, we consider only selfgravity
in the vertical direction, and assume that in the radial direction
centrifugal forces are balanced by gravity from a central mass ({\it 
Keplerian\/} approximation). In the following, we will often refer to such
accretion disks, in which selfgravity is important only in the vertical
direction, as {\it Keplerian selfgravitating (KSG) disks\/}.

Hydrostatic equilibrium in the vertical direction yields

\begin{equation}  \label{hydrostaticequilibrium}
P = \pi G \Sigma^2
\end{equation}
(Paczy\'nski 1978), where $P$ is the pressure in the central plane ($z = 0$), 
$\Sigma$ is the surface mass density integrated in the $z$ direction, and 
$G$ is the gravitational constant. Equation (\ref{hydrostaticequilibrium})
is strictly valid only in the limit of dominating selfgravity. We will
comment on the case of marginal selfgravity later. To deduce Eq.\ (\ref
{hydrostaticequilibrium}), we have made a one-zone approximation for the
disk's vertical structure, i.e., we have replaced derivatives with respect
to the $z$ direction by quotients of differences between the disk's surface
and the central plane.

Since details of the thermodynamics in the $z$ direction are of no
particular relevance to our argument, we shall likewise assume the disk to
be isothermal in the vertical direction.

Integrating the equation of conservation of angular momentum gives

\begin{equation}  \label{angularmomentum}
\nu \Sigma = - \frac{\dot M}{s^3 \omega^\prime}(s^2 \omega - \xi)
\end{equation}
with the radial mass flow rate\footnote{We choose the convention
of radial mass flow rate $\dot M$ and radial velocity $v_s$ {\it positive\/}
for inward motion.} $\dot M$, the Keplerian rotational frequency 
$\omega$, its radial derivative $\omega^\prime$, and a quantity $\xi$
allowing for the integration constant or, equivalently, for the inner
boundary condition. For a detailed discussion of $\xi$ see, e.g., Duschl
\&\ Tscharnuter (1991), Popham \&\ Narayan (1995) and Donea \&\ Biermann 
(1996). For simplicity, we set
the boundary condition $\xi = 0$ in the subsequent
discussion. This does not alter the
essence of our argument, and only changes details close to the disk's inner
radial boundary, since the product $s^2\omega$ increases with $s$.

Finally, we use the standard viscosity {\it ansatz\/}

\begin{equation}  \label{viscosity}
\nu = \alpha h c_{{\rm s}}
\end{equation}
and have for the sound velocity

\begin{equation}  \label{soundvelocity}
c_{{\rm s}}^2 = P \left/ \left( \frac{\Sigma}{2 h} \right) \right. .
\end{equation}
Equations (\ref{hydrostaticequilibrium}) and (\ref{soundvelocity}) give $h =
c_{{\rm s}}^2/(2 \pi G \Sigma)$. Substituting 
into Eq.\ (\ref{viscosity}), we get 
$\nu\Sigma = ( \alpha c_{{\rm s}}^3 ) / ( 2 \pi G ) $. Comparing this with Eq.
(\ref{angularmomentum}), we find that the set of Eqs. (\ref
{hydrostaticequilibrium}) - (\ref{soundvelocity}) can be fulfilled only for
a certain sound velocity, or temperature given by:

\begin{equation}  \label{constantcs}
c_{{\rm s}}^3 = - \frac{2 \pi G \dot M}{\alpha} \frac{\omega}{s \omega^\prime}
\sim - \frac{2 \pi G \dot M}{\alpha} \left(\frac{{\rm d} \log s}{{\rm d} \log
\omega}\right).
\end{equation}
and hence

\begin{equation}
c_{{\rm s}}^2=\frac{kT_{{\rm c}}}{m_H}=\left( \frac{2\pi }{3\alpha }G\dot{M}
\right) ^{2/3}.  \label{cssquare}
\end{equation}
Thus, for a KSG disk, the $\alpha $-ansatz leads to the requirement of a
constant temperature for all radii $s$ (or, if $\xi \not = 0$, the
temperature is prescribed as a function of $s$), independent of
thermodynamics. It also requires that the disk structure satisfy

\begin{equation}  \label{hsigma}
h \Sigma = \frac{c_{{\rm s}}^2}{2 \pi G} = \frac{\dot M^{2/3}}{\left( 3
\alpha \right)^{2/3} \left( 2 \pi G \right)^{1/3} }.
\end{equation}
In a standard non-SG accretion disk the temperature is a free parameter
which is determined by the energy released by the inward flow of the disk
gas ($\dot M$), by the local viscosity, and by the respective relevant
cooling mechanisms. The viscosity depends on $T_{{\rm c}}$ via Eq.\ (\ref
{viscosity}) and on the equation of hydrostatic support in the direction
normal to the disk, namely

\begin{equation}  \label{hydrostaticnsg}
\frac{h}{s} = \frac{c_{{\rm s}}}{v_\phi}
\end{equation}
($v_\phi = s \omega$) which in the non-SG case replaces Eq.\ (\ref
{hydrostaticequilibrium}).

In the KSG case, it is the surface density $\Sigma$ (and hence $h$) which
must adjust in order to radiate the energy deposited by viscous dissipation
and provided by the inward flowing material. While detailed solutions are
beyond the scope of this paper, they clearly exist formally. On the other
hand, the normal {\it thermostat\/} mechanism does not operate, at least in
the steady state. Indeed in certain circumstances, the condition of
constant mid-plane temperature appears to be inconsistent with the basic
thermodynamic requirement that the average gas temperature in the disk
exceed that of the black body temperature required to radiate away the
energy dissipated by viscous stresses (see Appendix).
We are therefore doubtful whether a physically plausible and stable 
quasi-steady state solution exists.

Furthermore, the Jeans condition for fragmentation in the disk into 
condensations of radius $R$ is

\begin{equation}  \label{jeans}
\frac{4\pi}{3} q G \overline{\rho} R^2 > c_{{\rm s}}^2
\end{equation}
(see Mestel 1965) where $q$ is factor of order unity. On the other hand,
Eqs.\ (\ref{hydrostaticequilibrium}) and (\ref{soundvelocity}) give

\begin{equation}  \label{pgsh}
2 \pi G \Sigma h = 4 \pi G \overline{\rho} h^2 = c_{{\rm s}}^2
\end{equation}
Thus, a selfgravitating disk is on the verge of fragmenting into
condensations of radius $R \sim h$ unless these are destroyed by shear
motion associated with the Keplerian velocity field.

Finally we note that the Reynolds number $\Re = v l / \nu_{{\rm mol}}$ in
the disk flow is extremely high in any astrophysical context and this in
itself is likely to lead to strong turbulence.

Indeed, as has been pointed out for example by Lynden-Bell and Pringle
(1974) and by
Thompson et al. (1977), the high Reynolds number character of the flow will
lead to the generation of turbulence and hence to a steady enhancement in
the effective (or Eddy) viscosity until the Reynolds number becomes
subcritical. The limiting viscosity in this case is given for a disk as

\begin{equation}  \label{nureynolds}
\nu = \frac{1}{\Re_{{\rm crit}}} v_\phi s = \beta v_\phi s
\end{equation}
where $\Re_{{\rm crit}} \sim 1/\beta \sim 10^2\ \dots\ 10^3$. This form is
quite different to the $\alpha$-ansatz.

For all of the above reasons, we are led to consider alternative
prescriptions for viscosity which may be more relevant to selfgravitating
Keplerian (or fully selfgravitating) disks.

\section{Prescription for Turbulent Viscosity\label{solutioneins}}

As noted in Sections \ref{introduction}\ and \ref{inconsistency}\ the need
for some kind of turbulent viscosity in accretion disks is generally
recognized, as is the very high Reynolds number of the flow in the absence
of such a viscosity. In any event, it seems reasonable to assume that the
turbulence is driven by the velocity field in the disk which has
characteristic length and velocity scales $s$ and $v_\phi$, respectively. We
might then expect that

\begin{equation}  \label{defnu}
\nu = v_{{\rm t}} l_{{\rm t}} \propto \Delta v_\phi \Delta s
\end{equation}
where $\Delta v_\phi$ and $\Delta s$ are representative velocity and length
scales of the flow. Furthermore, we may write

\begin{equation}  \label{deltav}
\Delta v_\phi = \frac{\partial v_\phi}{\partial s} \Delta s \sim 
\frac{v_\phi}{s} \Delta s
\end{equation}
If we consider turbulent elements in a {\it smoothed out\/} background gas
with sound speed $c_{{\rm s}}$, we may impose the limit that the turbulent
velocities do not exceed $c_{{\rm s}}$. Thus Eq.\ (\ref{deltav}) gives

\begin{equation}  \label{cslimit}
\frac{v_\phi}{s} \Delta s \sim \Delta v_\phi \le c_{{\rm s}}
\end{equation}
or

\begin{equation}
\Delta s \sim \frac{s}{v_\phi} \Delta v_\phi \le \frac{s}{v_\phi} c_{{\rm s}}
\end{equation}
From Eq.\ (\ref{hydrostaticnsg}), this implies that for a standard Keplerian
disk,

\begin{equation}  \label{deltas}
\Delta s \le h
\end{equation}
and hence that

\begin{equation}
\label{nuequ}
\nu \sim \alpha h c_{{\rm s}} 
\end{equation}
where $\alpha \le 1$. Note that the upper bound to $\Delta s$ {\it implies\/} 
approximately
isotropic turbulence. This is the standard $\alpha$-ansatz but derived from
considerations of rotationally generated turbulence. The argument can, in
fact, be inverted to show that under assumption of isotropic turbulence
Eqs.\ (\ref{hydrostaticnsg}) and (\ref{cslimit}), with $\Delta s = h$, imply 
$\Delta v_\phi \sim c_{{\rm s}}$ which again yields the functional
dependence of the $\alpha$-ansatz. Implicit in this argument is that the
effective viscosity is determined by the largest physically reasonable length 
and velocity scales.

For a SG disk (whether Keplerian or not) Eq.\ (\ref{hydrostaticnsg}) is no
longer valid so this analysis does not apply. In physical terms, the scale
height in the disk no longer reflects global properties of the disk (mass of
and distance to the central star) but is set by local conditions.
Furthermore, if the material is significantly clumped, as seems likely, for
example, in molecular clouds in galactic disks, an alternative
prescription may be more appropriate. We suggest that in such situations
(and possibly in others also) the appropriate length and velocity parameters
may be the radial distance, $s$, and the azimuthal velocity, $v_\phi$ or,
more specifically, some fractions of each, since in some sense they already
represent the maximum of either scale available.

In support of this choice of $s$ as the natural length scale we note that
{\it (a)\/} it is the only length scale which is relevant for angular momentum
transport which contains information about the driving agent for the
turbulence -- namely the rotation field, and {\it (b)\/} it is relevant 
to several
other processes for transporting angular momentum such as spiral arms and
large scale magnetic fields. Likewise, the orbital velocity is the only 
relevant velocity scale containing information on the rotation.

In these circumstances, we should obtain the {\it Rey\-nolds\/} form for the
viscosity, namely

\begin{equation}  \label{betaansatz}
\nu = \nu_{{\rm R}} = \beta s v_\phi
\end{equation}
where $\beta$ is a constant. Since the rotation field is assumed to be
driving the turbulence, it seems appropriate to restrict $\beta$ using the
condition that the effective Reynolds number of the flow not fall below the
critical value $\Re_{{\rm crit}}$ for the onset of turbulence. Thus $\beta$
should satisfy

\begin{equation}  \label{betalimit}
\beta < \frac{1}{\Re_{{\rm crit}}} \sim 10^{-3}\ \dots\ 10^{-2}
\end{equation}
In the following we will refer to Eq.\ (\ref{betaansatz}) as the 
$\beta$-{\it ansatz\/} and to the disk structure arising from this viscosity
prescription as $\beta$-{\it disks\/}. We suggest that this formulation is
most appropriate for selfgravitating disks which we explore further in
Sec.\ \ref{structure}.

We note, however, that if the disk matter distribution is clumpy (e.g.,
clouds within a low density smoothed out distribution) then there is a
formal connection between the $\alpha$- and the $\beta$-prescriptions. Since
in the $\beta$-formulation the clump velocities are of order $v_\phi$ shock
heating will tend to heat the low density interclump gas until its sound
speed $c_{{\rm s}} \sim v_\phi$. The interclump gas will then have a scale
height $h \sim s$, the scale of the clumpy disk, and will hence be roughly a
spherical structure. At this point the $\alpha$- and $\beta$-prescriptions
look formally identical but the scale height and sound speed now refer to a
more or less spherical background distribution of hot gas in which a
disk structure of cloudy clumps is inbedded. 

\section{Structure of Selfgravitating ${\bf \beta }$ Disks\label{structure}}

For SG $\beta$-disks, Eqs.\ (\ref{hydrostaticequilibrium}), (\ref
{angularmomentum}) and (\ref{soundvelocity}) are applicable, but Eq.\ (\ref
{viscosity}) must be replaced by Eq.\ (\ref{betaansatz}). It then follows
that

\begin{equation}  \label{sigmabeta}
\Sigma = - \frac{\dot M}{\beta s^3 \omega^\prime}
\end{equation}
and that

\begin{equation}
c_{{\rm s}}^2 = - \frac{\pi G h}{\nu} \frac{\dot M}{s \omega^\prime} = 
- \frac{ \pi G h \dot M}{\beta s^3 \omega^\prime}
\end{equation}
Thus the SG $\beta$-disks recover the thermostat property of the standard
disk, namely that the temperature and scale height can adjust to accomodate
(radiate away) the energy input to the system from viscous dissipation and
inward motion.

\subsection{Keplerian Selfgravitating Disks}

For the particular case of a KSG $\beta$-disk we have:

\begin{equation}
\frac{c_{{\rm s}}^2}{h} = \frac{4}{3} \frac{\pi G \dot M}{\beta} \frac{1}{(G
M_*)^{1/2} s^{1/2}}
\end{equation}
For the SG $\beta$-disk it follows immediately from Eq.\ (\ref{sigmabeta})\
and from mass conservation in the disk that the radial inflow velocity $v_s$
is given by

\begin{equation}  \label{vssg}
v_s = \frac{\dot M}{2 \pi s \Sigma} = - \frac{\beta s^2 \omega^\prime}{2 \pi}
\end{equation}
For the KSG $\beta$-disk, we then have

\begin{equation}  \label{vsksg}
v_s = \frac{3}{4 \pi} \beta s \omega = \frac{3}{4 \pi} \beta v_\phi
\end{equation}
Thus at each radius the inward velocity is the same fraction of the local
orbital velocity. From Eq.\ (\ref{vssg}) this, in fact, holds for any SG 
$\beta$-disk in which the angular velocity is a power law function of $s$
with adjustment only to the numerical factor in Eq.\ (\ref{vsksg}). If 
$\beta $ satisfies the constraint (\ref{betalimit}), then the approximation of
centrifugal balance in the radial direction remains well justified.

Under these conditions the dissipation per unit area of a SG $\beta$-disk is
given by

\begin{equation}  \label{dissipation}
D = \frac{\dot M}{4 \pi s} \left(\frac{v_\phi^2}{s}\right) = 2 \sigma 
T_{{\rm eff}}^4
\end{equation}
where $\sigma$ is the Stefan-Boltzmann constant.
For an optically thick KSG disk this yields the same radial dependence of 
$T_{{\rm eff}}$ as for the standard disk, namely

\begin{equation}  \label{teff}
T_{{\rm eff}} = \left(\frac{G \dot M M_*}{8 \pi \sigma}\right)^{1/4} s^{-3/4}
\end{equation}
This temperature dependence which is identical to that of the standard model
then leads to the well known energy distribution for an optically thick
standard disk of $S_\nu \propto \nu^{1/3}$. This also implies that -- as
long as the disks are not fully self-gravitating -- it is hard to distinguish
between an $\alpha$- and a $\beta$-disk model observationally. 

\subsection{Fully Selfgravitating Disks (FSG)}

We turn now to the case of the fully selfgravitating (FSG) $\beta$-disk, in
which the disk mass is sufficiently great that it dominates the
gravitational terms in the hydrostatic support equation in both the radial
and vertical directions. While there are many potential solutions for the
FSG disk structure, one is well known in both mathematical and observational
terms, namely the constant velocity ($v_\phi = {\rm const.}$) disk. Within
such a disk structure we have simultaneous solutions to the equation of
radial hydrostatic equilibrium and Poisson's equation of the form

\begin{equation}  \label{vphisigma}
%v_\phi = s \omega = v_0\hspace{1cm}{\rm and}\hspace{1cm} \Sigma \propto
v_\phi = s \omega = v_0\hspace{4mm}{\rm and}\hspace{4mm} \Sigma \propto
\Sigma_0 \left(\frac{s}{s_0}\right)^{-1}.
\end{equation}
(Toomre 1963; Mestel 1963). For the FSG disk, Eq.\ (\ref{sigmabeta}) then
leads to

\begin{equation}  \label{sigmafsg}
\Sigma = \frac{\dot M}{\beta s v_0}
\end{equation}
which has the same radial dependence as the structural solution as shown in
Eq.\ (\ref{vphisigma}). Thus Eq.\ (\ref{sigmafsg}) may be viewed as giving
the rate of mass flow through the disk for a FSG $\beta$-disk with constant
rotational velocity $v_0$. Finally, the equation of continuity provides a
constraint if the structure is to maintain a basically steady state
structure. We then have:

\begin{equation}
s\frac{\partial \Sigma}{\partial t} = \frac{\partial}{\partial s}(s v_s
\Sigma) = \frac{3}{4 \pi} \frac{\partial}{\partial s}(\beta v_\phi s \Sigma)
\end{equation}
For the $v_\phi = v_0 = {\rm const.}$ disk, it then follows from Eq.\ (\ref
{sigmafsg}) that $\partial\Sigma / \partial t = 0$. Thus the constant
velocity disk represents a steady state solution in regions sufficiently far
from the inner and outer boundaries of the $\beta$-disk.

It is then possible, in the spirit of the discussion of Eqs.\ (\ref
{dissipation}) and (\ref{teff}), to calculate the energy dissipation rate
per unit area $D$ for the constant velocity $\beta$-disk. We then find

\begin{equation}
D = 2 \sigma T_{{\rm eff}}^4 = \frac{\dot M v_0^2}{4 \pi s^2}
\end{equation}
so that

\begin{equation}
T_{{\rm eff}} = \left(\frac{\dot M v_0^2}{8 \pi \sigma} \right)^{1/4} s^{-1/2}
\end{equation}
For flux density $S_\nu$ emitted by an optically thick, constant velocity 
$\beta$-disk it then follows that

\begin{equation}
S_\nu \propto \nu^{-1}
\end{equation}
In reality, a sufficiently massive disk may be expected to have an inner
Keplerian (standard) zone, a Keplerian selfgravitating zone (KSG), and a
fully selfgravitating zo\-ne (FSG). We should therefore expect a smooth
transition in the spectral energy distribution from the $\nu^{1/3}$ spectrum of
the inner two zones to the $\nu^{-1}$ spectrum arising at longer wavelengths
from the FSG zone. One could turn this argument around and argue that, if no
other components contribute to the spectrum, the flatness of the $\nu F_\nu$
distribution is a measure for the importance of selfgravity and thus for the
relative mass of the accretion disk as compared to the central accreting
object. This, of course, applies only to the optically thick case which may
not arise frequently in strongly clumped disks.

\subsection{Time Scales\label{secttimescales}}

The evolution of accretion disks can be described by a set of time scales.
For our purposes, the dynamical and the viscous time scale are of particular
interest.

The dynamical time scale $\tau_{{\rm dyn}}$ is given by

\begin{equation}
\tau_{{\rm dyn}} = \frac{1}{\omega}.
\end{equation}
While this formulation applies to all cases, selfgravitating or not, 
it is only in the non-SG and in the KSG cases that $\omega$ is given
by the mass of the central accretor and by the radius. In the FSG case, 
$\omega$ is determined by solving Poisson's equation.

The timescale of viscous evolution $\tau_{{\rm visc}}$ is given by

\begin{equation}
\label{equtauvisc}
\tau_{{\rm visc}} = \frac{s^2}{\nu}
\end{equation}
In the standard non-SG and geometrically thin ($h \ll s$) case, this leads to

\begin{equation}  \label{tauviscnsg}
\tau_{{\rm visc}}^{{\rm non-SG}} = \left(\frac{s}{h}\right)^2 
\frac{\tau_{{\rm dyn}}}{\alpha} \gg \frac{\tau_{{\rm dyn}}}{\alpha}.
\end{equation}
In KSG and FSG disks, $\tau_{{\rm visc}}$ is given by

\begin{equation}  \label{tauviscsg}
\tau_{{\rm visc}}^{{\rm KSG}} = \tau_{{\rm visc}}^{{\rm FSG}} = \tau_{{\rm 
visc}}^{{\rm SG}} = \frac{\tau_{{\rm dyn}}}{\beta}
\end{equation}
With $\alpha < 1$ and $\beta \ll 1$ (Eq.\ \ref{betalimit}) under all
circumstances $\tau_{{\rm visc}} \gg \tau_{{\rm dyn}}$. In the SG cases
the ratio between the two timescales decouples from the disk structure.
In all cases the models are self consistent in assuming basic hydrostatic
equilibrium in the vertical direction.

\section{Possible Applications\label{examples}}

\subsection{Protoplanetary Accretion Disks}

T Tauri stars have infrared spectral energy distributions $\nu F_\nu$ which
can be approximated in many cases by power laws $\nu F_\nu \propto \nu^n$
with a spectral index $n$ in the range $\sim 0 \dots 1.3$. Assuming this
spectral behaviour to be due to radiation from an optically thick disk, it
translates into a radial temperature distribution $T_{{\rm eff}} \propto
s^{-q}$ with $n = 4 - 2/q$.

A {\it passive\/} disk that only reradiates reprocessed radiation from the
central star has $q = 3/4$ or $n = 4/3$. An optically thick {\it active\/}
non-selfgravtating accretion disk which radiates energy that is liberated
through viscous dissipation shows the same spectral distribution. Thus, if
accretion disks were the only contributors to T Tauri spectra, and if the
disks were optically thick and non-selfgravitating, the spectral shape
should be the same for all objects. This is in clear contrast to the above
mentioned observed spectral energy distributions.

Adams, Lada and Shu (1988) were the first to discuss the possibility of a
non-standard radial temperature distribution with $q \not = 3/4$. Using $q$
as a free parameter, they find that for flat spectrum sources, their best
fits require disk masses that are no longer very small as compared to the
accreting stars' masses. They already mention the possibility that the
flatness of the spectrum and selfgravity of the disk may be related. On the
other hand, at that time this indirect argument was the only evidence for
large disk masses.

Due to the lack of accretion disk models yielding the appropriate values of 
$q$, since the work of Adams, Lada and Shu several other interpretations for 
the flatness of the spectra were offered, for instance dusty envelopes 
engulfing a star with a standard disk around it (Natta 1993).

In the meantime, high resolution direct observations of protostellar disks
yield independent strong evidence for comparatively large disk masses. Lay
et al.\ (1994), for instance, find a lower limit for the disk masses in HL
Tau -- one of the sources in Adams, Lada and Shu sample of flat spectrum T 
Tauri stars -- of $\sim 0.02\,{\rm M}_\odot$.

We suggest that the flatness of the spectrum actually reflects the mass of
the disk, i.e., the importance of selfgravity.
For disk masses considerably smaller than $\sim 1/30M_{*}$, the standard
accretion disk models apply. For disks whose masses are larger but still
small compared to $M_{*}$ the spectral behaviour is not altered significantly.
but disk structure and the time scale of disk evolution ($\tau _{{\rm 
visc}}$, see Eqs.\ \ref{tauviscnsg}\ and \ref{tauviscsg}) change. For even more
massive disks, we expect a clear trend towards flatter spectra that approach
an almost constant $\nu F_\nu $ distribution if selfgravity in the disk
becomes important in radial as well as in vertical direction. But the time
scale for the disk evolution remains the same (Eq.\ \ref{tauviscsg}).

\subsection{Galactic Disks}

The relevance of viscosity  in the evolution of galactic disks has been the 
subject of discussion since von Weizs\"acker (1943, 1951) and L\"ust
(1952) first raised the issue nearly fifty years ago. They noted then that, 
with an eddy viscosity formulation (a $\beta$-disk), 
the time scale for evolution of typical galactic disks was 
comparable to the age of the universe and suggested that this might account 
for the difference between spiral and elliptical galaxies. 

With the subsequent realization that galactic disks moved primarily under the 
influence of extended massive halos, interest in FSG disks waned. However, as 
noted above, it is possible for a massive disk to  exist and evolve under the 
influence of viscosity embedded in such a halo gravitational field. Indeed, 
in the event that such a structure forms, it must evolve under viscous 
dissipation and can achieve a quasi-steady state with essentially the same 
mass  and energy dissipation distribution as for the FSG constant velocity 
disk. We refer to this case as an Embedded Self Gravitating (ESG) disk.

The timescale for viscous evolution $\tau_{\rm visc}$ as given in section 
\ref{secttimescales}\ suggests a means of differentiating between the 
$\alpha$- and $\beta$-formulations for this case. For a normal spiral galaxy 
with a 
suggested mean temperature in the gaseous disk of around $10^4\,$K and a scale 
height of around 300\, pc, we obtain

\begin{eqnarray}
\tau_{\rm visc}^{(\alpha)} & \sim & 10^4 \tau_{\rm dyn} \sim 3\,10^{11}\,{\rm 
yr}\nonumber\\
\tau_{\rm visc}^{(\beta)} & \sim & 10^2 - 10^3 \tau_{\rm dyn} \sim 3\,10^9 - 
3\,10^{10}\,{\rm yr}
\end{eqnarray}

Thus, with these parameters, little evolution would take place in a Hubble 
time on the $\alpha$-hypothesis but significant evolution is predicted on the 
$\beta$-hypothesis.

In terms of inflow velocities the $\beta$-ansatz suggests values in the range 
$0.3 - 3\,$km\,s$^{-1}$ which would be exceedingly hard to measure directly: 
the $\alpha$-ansatz suggests still lower values. On the other hand, it may be 
possible to provide limits on the viscosity through other observational 
constraints. For example, the build up of the 3\,kpc molecular ring in our own 
galaxy can be interpreted as due to viscosity driven inflow in the constant 
velocity part of the galactic disk which ceases (or at least slows down) in 
the constant angular velocity inner regions (Icke 1979; D\"ather and Biermann
1990). Similarly, several authors 
have suggested that the radial abundance gradients observed in our own and 
other disk galaxies may be due to radial motion and diffusive mixing 
associated with the turbulence generating the eddy viscosity (Lacey \&\
Fall 1985; Sommer-Larsen and Yoshii 1990; K\"oppen 1994; Edmunds \&\ Greenhow 
1995, Tsujimoto, Yoshii, Nomoto and Shigeyama 1995). 
According to these authors, radial inflows of around 
1\,km\,s$^{-1}$ at the galactic location of the Sun are required for optimum 
fits to the abundance gradient data within the context of the viscous disk 
hypothesis. Such inflow velocities are consistent with the $\beta$-ansatz but 
could, of course, be generated also by other means (e.g effects of bars, 
magnetic fields).

\section{Summary}
The standard model for geometrically thin accretion disks with viscosity 
proportional to sound velocity and vertical scale height (often 
referred to as $\alpha${\it -disks\/}) leads to inconsistencies if the disk's
mass is large enough for selfgravity to play a role. This problem arises even
in Keplerian selfgravitating disks in which only the vertical structure is 
dominated by selfgravity while the azimuthal motion remains Keplerian.

We propose a viscosity prescription based on the assumption that the 
effective Reynolds number of the turbulence does not fall below the critical 
Reynolds number. In this parametrization the viscosity is proportional to the 
azimuthal velocity and the radius ($\beta$-disks). This prescription yields
physically consistent models of both Keplerian and fully selfgravitating 
accretion disks. Moreover, for the case of sufficiently small disk masses, 
we recover the $\alpha$-disk solution as a limiting case.

Such $\beta$-disk models may be relevant to protoplanetary accretion disks 
as well as for galactic disks. In the case of protoplanetary disks they yield 
spectra that are considerably flatter than those due to non-selfgravitating 
disks, in better agreement with observed spectra of these objects. In galactic 
disks, they result in viscous evolution on time scales shorter than the Hubble 
time and thus offer a natural explanation for an inward flow that could
account for the observed chemical abundance gradients.

\acknowledgments
We thank Achim Traut, Heidelberg, for helpful comments on the manuscript. 
WJD acknowledges partial support by the {\it Deutsche Forschungsgemeinschaft 
DFG\/} through {\it SFB 328 (Evolution of Galaxies)\/}.

\appendix

\section{Thermodynamic Considerations for KSG $\alpha$-disks}

For a KSG $\alpha$-disk, we have from Eq.\ \ref{cssquare}\ that

\begin{equation}
T_{\rm c} = \left( \frac{2\pi}{3\alpha} G \dot M \right)^{2/3} 
\frac{m_{\rm H}}{k} = 2.41\,10^5\,{\rm K}\,\left(\frac{\dot m}{\alpha_{-1}}
\right)^{2/3}
\end{equation}
where $\dot m$ is the mass flow rate in solar masses per year, and 
$\alpha_{-1} = \alpha/0.1$.

If the disk is optically thick and advection is negligible, viscous 
dissipation leads to local effective temperature of

\begin{eqnarray}
T_{\rm eff} & = & \left( \frac{3}{8\pi\sigma} \right) \left( G M \right)^{1/4}
\dot M^{1/4} s^{-3/4}\nonumber\\
& = & 8.53\,10^3\,{\rm K} \left(\frac{m \dot m}{s_{\rm
A}^3}\right)^{1/4}
\end{eqnarray}
with $m$ the mass of the central star in solar units and $s_{\rm A}$ the
radius in astronomical units.

An essential thermodynamics requirement is that $T_{\rm c} > T_{\rm eff}$ or
that

\begin{equation}
\frac{T_{\rm eff}}{T_{\rm c}} = 3.53\,10^{-2} \frac{m^{1/4} \dot 
m^{-5/12}}{s_{\rm A}^{3/4}} \alpha_{-1}^{2/3} < 1
\end{equation}
This condition is satisfied provided that 

\begin{equation}
\dot m > \dot m_{\rm T} = 3.27\,10^{-4} \frac{m^{3/5} 
\alpha^{8/5}_{-1}}{s_{\rm A}^{9/5}}
\end{equation}
and that the disk is selfgravitating in the vertical direction at $s_{\rm A}$.
The latter condition leads to a second requirement on $\dot m$.

For a standard Keplerian disk, the mass flow rate is given (Eqs.\
\ref{hydrostaticnsg},\ref{nuequ},\ref{equtauvisc}) by

\begin{equation}
\label{appequmdot}
\dot M \approx \frac{M_{\rm d}}{\tau_{\rm visc}} \approx \frac{M_{\rm d} 
\nu}{s^2} = \alpha M_{\rm d} \left(\frac{h}{s}\right)^2 \omega
\end{equation}
with $M_{\rm d}$ the disk's mass.
From Eq.\ \ref{eqcondvertsg}, the condition that the disk is 
non-selfgravitating is $M_{\rm d} < (h/2s) M_*$ and hence, from Eq.\ 
\ref{appequmdot}, that

\begin{equation}
\dot M < \frac{\alpha}{2} \left(\frac{h}{s}\right)^3 \left(\frac{G M_*^3}{s^3}
\right)^{1/2}
\end{equation}
or

\begin{equation}
\dot m < \dot m_{\rm G} = 3.14\,10^{-1} \alpha_{-1} \left(\frac{h}{s}\right)^3
\frac{m^{3/2}}{s_{\rm A}^{3/2}}
\end{equation}

A selfconsistent and physically acceptable solution can be obtained only if
$\dot m_{\rm G} > \dot m_{\rm T}$, that is the disk becomes selfgravitating at 
values of $\dot m$ which are sufficiently high that thermodynamic 
requirements are not violated. This condition may then be written
as 

\begin{equation}
\frac{h}{s} > 1.01\,10^{-1} m^{-3/10} s_{\rm A}^{-1/10} \alpha_{-1}^{1/5}
\end{equation}
Thus {\it thin\/} KSG $\alpha$-disks with $m \not \gg 1$, $s_{\rm A} \not \gg 
1$ appear to be inconsistent with basic thermodynamic requirements. 

No such inconsistencies occur for $\beta$-disks for which the normal 
thermostat effect is free to operate.

\end{document}